# An Efficient Word Lookup System by using Improved Trie Algorithm

Rahat Yeasin Emon
Department of Computer Science and Engineering
Chittagong University of Engineering and Technology
Chittagong-4349, Bangladesh
emon13111@gmail.com

Sharmistha Chanda Tista
Department of Computer Science and Engineering
Chittagong University of Engineering and Technology
Chittagong-4349, Bangladesh
tistacuet09@gmail.com

*Abstract*—**Efficiently word storing and searching is an important task in computer science. An application's space complexity, time complexity, and overall performance depend on this string data. Many word searching data structures and algorithms exist in the current world but few of them have space compress ability. Trie is a popular data structure for word searching for its linear searching capability. It is the basic and important part of various computer applications such as information retrieval, natural language processing, database system, compiler, and computer network. But currently, the available version of trie tree cannot be used widely because of its high memory requirement. This paper proposes a new Radix trie based data structure for word storing and searching which can share not only just prefix but also infix and suffix and thus reduces memory requirement. We propose a new emptiness property to Radix trie. Proposed trie has character cell reduction capability and it can dramatically reduce any application's runtime memory size. Using it as data tank to an operating system the overall main memory requirement of a device can be reduced to a large extent.**

## I. INTRODUCTION

Word lookup algorithm is very important in modern computer science. A software application's overall performance depends on this word searching algorithm. Nowadays world is demanding new data lookup, data structure and algorithm to improve software performance. In that case, the proposed methodology will develop a new word lookup data structure which has unique character cell reduction ability.

Trie is an awesome data structure where searching time complexity is $O(L)$, where '$L$' is the length of the searched word. In the case of searching time complexity, no other algorithm and data structure can beat trie tree. But the problem is that it requires a huge amount of memory to build a trie tree. Trie tree's most modern variant Radix trie can reduce memory consumption a little bit. Radix trie can only share prefix data it does not have infix and suffix sharing ability. We focus on Radix trie. We have improved this Radix trie and claimed that in the case of searching time complexity and space complexity our proposed trie tree is better than Radix trie and some other popular data structures that are currently used. We will show a comparison between our improved trie tree and some most common data structures. We will also show the data tank properties of proposed trie.

## II. BACKGROUND

Trie tree is a character wise tree. For improvement of space complexity of trie tree, there exists some theoretical and practical work. Trie tree structure first proposed by de la Briandais (1959) [1]. In this paper, they proposed array and linked list based implementation of child list and one node contain only one character. Lots of unused nodes were created in this trie structure. For improving space complexity of native trie tree Donald R. Morrison invented a compressed trie called Radix trie in 1968 [2]. The main improvement was that this algorithm merged nodes when the nodes in the native trie tree form a single character chain. In 2002 Heinz, Zobel & Williams proposed Burst trie [3]. It was an improved version of Radix trie. They followed the same structure of Radix trie and used two or three levels of trie tree or some other data structures to reduce the overall structure of Radix trie. It is still memory inefficient not usable in most cases. Nikolas Askitis and Ranjan Sinha proposed HAT trie in 2007 [4]. This HAT trie is a variant of Burst trie which carefully combines the combination of data structures that are used in Burst trie structure. All the trie tree variants only have prefix sharing capability. In 2000 Jan Daciuk, Stoyan Mihov, Bruce Watson, and Richard Watson proposed deterministic acyclic finite state automaton (DAFSA) [5] which have a prefix and suffix sharing ability. But the problem was that for searching similar suffix it has to traverse the full DAFSA tree which was very time inefficient. DAFSA could not directly store auxiliary information relating to each path and cannot have infix sharing ability. Kurt Maly, Univ. of Minnesota and Minneapolis proposed C-Trie [6]. This methodology was a slight modification of the trie tree. Dan E.Willard proposed x-fast trie and y-fast trie in 1982 [7]. This methodology was quite similar to native Radix trie. Peter Gjol Jensen, Kim Guldstrand Larsen, and Jiri Srba proposed another Radix trie variant PTrie (Prefix-Trie) in 2017 [8]. To remove an unused node in trie tree Bentley and Sedgewick introduced the "ternary trie" in 1997 [9]. Its space complexity is similar to Radix trie but lookup time is $O(\log n + length(q))$, where '$n$' is the number of word exits in the ternary trie tree and '$q$' is the searched word. All of the trie tree variants above can only compress or share prefix data. Here we proposed a trie tree which has prefix, infix and suffix sharing ability.

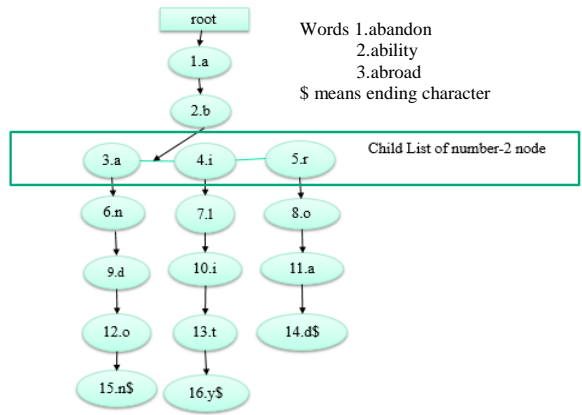

Fig. 1. Trie tree

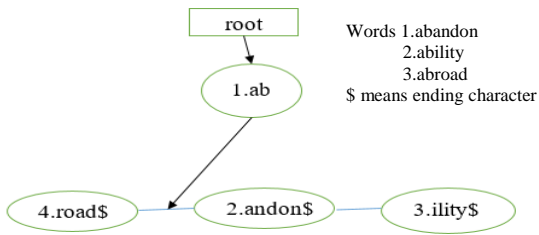

Fig. 2. Radix trie

### III. PROPOSED TRIE

In our proposed trie tree, we are introducing a new property to Radix trie. In Radix trie every node contains some string data and there are lots of nodes which have the same data. Why we store the same type of data or nodes multiple times? This is an unintelligent way and it requires a huge amount of memory to represent a Radix trie. We are focusing on this problem. We try to empty the Radix trie nodes as much as possible. To do this, unlike Radix trie after creating new node we do not put data directly to the node. We treated this entry data as a new word and insert it to the trie tree. By recursively doing this step we have found very compress radix trie tree and experimentally we have found that most of the trie nodes are empty, the requirement of character cell is very less and thus the memory requirement to represent this trie is also very low. Now, we will briefly explain the total algorithm. To empty trie tree nodes, we do not put data directly to the node rather we again insert the entry node data to the trie tree. We first check whether there exists a valid character path from root node which is equivalent to the entry data. Here character path means the character sequence needs to reach a node from the root. If a character path exists then it just points the last node of the character path as a data node. If not then it creates the desired character path and points the last node of the character path as a data node. With this trick, we can easily retrieve the desired data by traversing the tree upward. If we can only find or create valid character path for some of the prefix of entry data, then point the last node as a data node and rest of the suffix data put directly to the node. Else if we do not find a valid character path and we can not create a valid character path then only enter data directly to that node.

### A. Build the proposed Trie tree

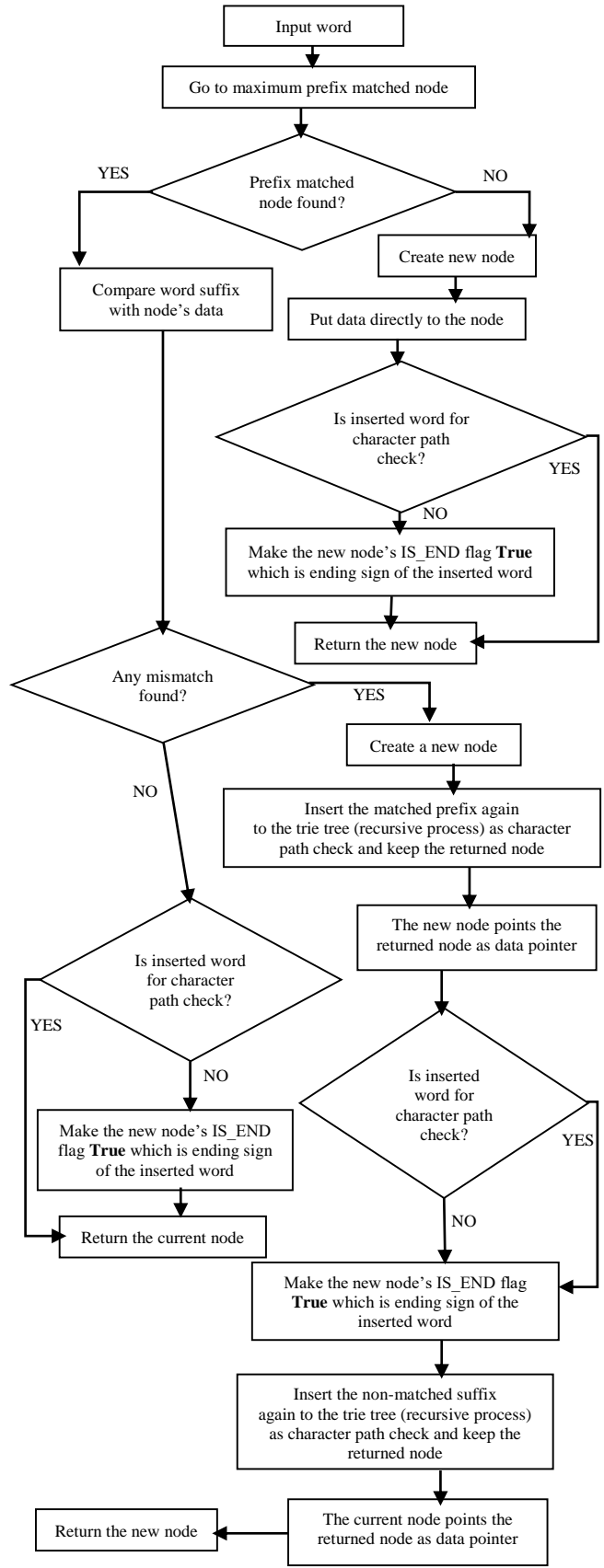

Fig. 3. Flowchart of the insertion algorithm

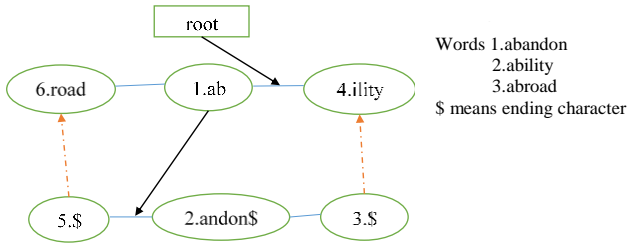
Words 1.abandon
2.ability
3.abroad
$ means ending character

Fig. 4. Proposed trie

The above figure shows the structure of proposed trie tree. Here, number-6, number-1, and number-4 are sibling nodes they are at the same level. Number-5, number-2, and number-3 are another sibling level. Our main goal is to create empty node as much as possible. Here, number-5 and number-3 node are empty nodes. They do not need to store data. Number-5 node can inherit data from the number-6 node and number-3 node can inherit data from the number-4 node. The proposed data structure needs less amount of character cells compare to native trie and radix trie tree.

*B. How proposed Trie remove data redundancy of Radix Trie*

Suppose, we want to insert '*abandon*' to Radix trie and the proposed trie. The corresponding tree looks.

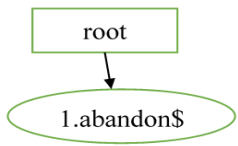    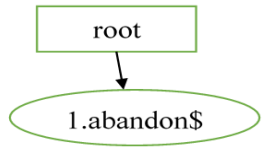

Fig. 5. Radix trie          Fig. 6. Proposed trie

Next, we insert '*ability*'. Then the corresponding trie becomes.

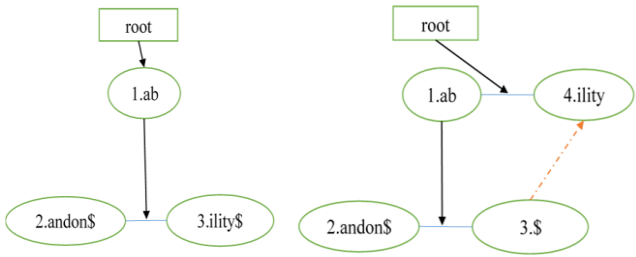

Fig. 7. Radix trie          Fig. 8. Proposed trie

In proposed trie, for removing data redundancy no node stores the same data internally. For doing that, we check or create a character path from the root. Here we create a character path '*ility*' (number-4 node) from the root and points number-3 node's data pointer to the number-4 node. Why this trick is used or helpful? This trick is because, when inserting further words to trie tree, there might have some node with same data '*ility*'. With this trick, we can easily find if there exists any node or character path with data '*ility*'. If there exists, then just point the data pointer to that corresponding node. One example clarifies this. Suppose in the insertion process our next word is '*abandonility*'. Now, look how our trie looks like.

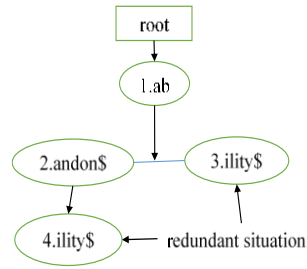    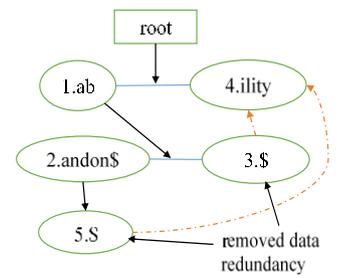

Fig. 9. Radix Trie          Fig. 10. Proposed trie

Look how proposed trie reduces data redundancy. In Radix trie figure number-3 and number-4 node have the same data '*ility*'. But in the proposed trie, we create '*ility*' (number-4 node) character path from root and number-3 and number-5 the two empty nodes just point number-4 node as the data pointer. In our next insertion process, there might have hundreds node with same data '*ility*'. This is the main fault of Radix trie. But proposed trie removes this fault. In the above example, we create a character path for '*ility*' and all other nodes just point the last node of '*ility*' character path (here number-4 node). With this trick, in the experimental analysis section, we will see maximum nodes are empty and the requirement of the character cell is very low.

*C. Implementation Strategy of Proposed Trie*

We use the AVL tree for child list representation. For this strategy, there is no character limitation of our trie tree, and searching time is also less. For storing data in one node, we use a linked list.

IV. EXPERIMENTAL ANALYSIS AND COMPARISON

In this section, we experimentally analyze the proposed trie and compare it with radix trie and some other popular data structures and algorithms. We use the low-level memory consumption approach to compare the algorithms.

*A. Character cell reduction of Proposed Trie*

The proposed trie need not store the same character sequences more than once. The emptiness property reduces the character cell requirement. Now we will show the character cell requirement of proposed trie for different set of input.

Table 1: Proposed trie character cell reduction

| Input | Description | Total Number of Character cells | Number of Character cells need to build the Improved trie |
|---|---|---|---|
| 20,000 english word [10] | Most common english word | 4,72,145 | 33 |
| 4,66,544 english word [11] | All available english word | 43,96,422 | 130 |
| 9,33,083 english word | Adding '1' to every word of second row | 92,59,388 | 130 |
| 13,99,620 english word | Adding '1' and '2' to every word of second row | 1,41,22,354 | 130 |

20,000 english words which have total 4,72,145 character cells. The proposed trie needs only 33 character cells to represent this huge data set in a tree structure format. For representing 4,66,544 english words it needs 130 character cells. It requires 8 bits to represent an ASCII character cell in memory and 16, 32 or more number of bits to represent a Unicode character cell. So character cell is the main focusing point to design a memory efficient data structure. From the above table, we see that proposed trie needs a very small amount of character cells to represent any huge word data set. From the above analysis, we can say that proposed trie is extremely memory efficient data structure.

*B. Radix Trie vs Proposed Trie*

Here, we will show an implementation analysis between radix trie and the proposed trie tree. We use the same implementation strategy (AVL tree - linked list) to implement radix trie and the proposed trie.

Table 2: Radix trie vs Proposed trie

| Input | Number of Character Cell required to Store this words | Number of AVL Node | | Number of Character Node | | Difference (Power of Our Improved Trie) |
|---|---|---|---|---|---|---|
| | | Radix Trie | Improved Trie Tree | Radix Trie | Improved Trie Tree | |
| 20,000 english word | 4,72,145 | 23,525 (All are non-empty) | 25,922 (25,894 are empty) | 47,363 | 33 | Huge |
| 4,66,544 english word | 43,96,422 | 6,08,368 (All are non-empty) | 6,46,912 (6,46,820 are empty) | 14,20,649 | 130 | Huge |
| 9,33,083 english word | 92,59,388 | 10,74,907 (All are non-empty) | 11,13,450 (11,13,358 are empty) | 18,87,187 | 130 | Huge |
| 13,99,620 english word | 1,41,22,354 | 15,41,444 (All are non-empty) | 15,79,983 (15,79,891 are empty) | 23,53,720 | 130 | Huge |

See how the proposed trie tree compresses data. For storing 20,000 english words radix trie needs 23,525 nonempty AVL nodes and 47,363 character cells. For storing the same data proposed trie needs 25,922 AVL nodes on them 25,894 nodes are empty nodes. That means only 92 (25,922 - 25,894) AVL node needs to store data directly. It needs only 33 character cells to represent this huge data. For 4,66,544 english words, the proposed data structure needs 130 character cells. From the above table, we see that the difference of character cell requirement between two data structures is huge. The emptiness property of proposed trie reduces the character cell requirement. From the above analysis, we can say that the proposed trie is extremely memory efficient than radix trie.

*C. Radix Trie vs Proposed Trie (Memory Consumption)*

Now, will analyze the low-level memory requirement of radix trie and the proposed trie. We use windows task manager to figure out the memory consumption.

Table 3: Radix trie vs Proposed trie (Memory Consumption)

| Input | Description | Memory Consumption Radix Trie | Memory Consumption Improved Trie | Difference/Improved Amount of Memory |
|---|---|---|---|---|
| 20,000 english word | Most common english word | 2,228 KB | 1,520 KB | 708 KB |
| 4,66,544 english word | All available english word | 46,824 KB | 25,948 KB | 20,876 KB |
| 9,33,083 english word | Adding '1' to every word of second row | 72,532 KB | 44,236 KB | 28,296 KB |

For storing 20,000 english words in radix trie consumes 2,228 KB memory space and the proposed trie needs only 1,520 KB memory space. When the data size is 4,66,544 english words the deference of memory consumption is 20,876 KB. From the above table, we see that when the input data size increased the difference of memory consumption is also increased dramatically. From the low-level analysis, we can say that proposed trie is memory efficient compared to radix trie.

*D. Native Trie vs Proposed Trie*

Though native trie is memory inefficient it is also used widely because of its design simplicity. Here, we will show an experimental analysis between native trie and the proposed trie tree. We use the most common array based child list representation to implement native trie. We use windows task manager to figure out the memory consumption.

Table 4: Native trie vs Proposed trie

| Input | Description | Memory Consumption Native Trie | Memory Consumption Improved Trie | Difference/Improved Amount of Memory |
|---|---|---|---|---|
| 20,000 english word | Most common english word | 6,124 KB | 1,520 KB | Huge |
| 4,16,290 english word | All available english word | 1,32,812 KB | 22,232 KB | Huge |

From the above table, we see that the for storing 20,000 english words in native trie tree it requires 6,124 KB memory space and the proposed trie needs only 1,520 KB memory space. When the input data size increased the difference between the memory consumption is also increased. The proposed trie is extremely memory efficient than native trie.

*E. AVL Tree vs Proposed Trie*

AVL tree and Red-Black tree are the two popular binary search tree. They are important data structures for efficiently retrieving data. Their searching time complexity is $O(log\ n)$ ('*n*' is the number of word) but they don't have any space compress ability. They have the same design structure. Here we will show a memory consumption analysis between our proposed trie and AVL tree. We use windows task manager to figure out the memory consumption.

Table 5: AVL tree vs proposed trie (Memory Consumption)

| Input | Description | Memory Consumption AVL_Tree | Memory Consumption Improved Trie | Difference/Improved Amount of Memory |
|---|---|---|---|---|
| 20,000 english word | Most common english word | 1,632 KB | 1,520 KB | 112 KB |
| 4,66,544 english word | All available english word | 27,064 KB | 25,948 KB | 1,116 KB |
| 9,33,083 english word | Adding '1' to every word of second row | 58,248 KB | 44,236 KB | 14,012 KB |

To store 20,000 english words AVL tree consumes 1,632 KB of memory space, on the other hand, proposed trie needs 1,520 KB of memory. When the input data size increased the difference of memory consumption is also increased dramatically. From the above table, we can say that proposed trie is memory efficient than AVL tree.

*F. C++ STL vs proposed Trie*

In our current software development, C++-built-in STL Map, Java-HashMap and Python-Data Dictionary are used for string data storing and searching. Here we see a comparison between the proposed trie and C++ STL Map.

Table 6: C++ STL vs proposed trie

| Input | Description | Memory Consumption C++ STL Map | Memory Consumption Improved Trie | Difference/Improved Amount of Memory |
|---|---|---|---|---|
| 20,000 english word | Most common english word | 1,720 KB | 1,520 KB | 200 KB |
| 4,66,544 english word | All available english word | 30,744 KB | 25,948 KB | 4,796 KB |
| 9,33,083 english word | Adding '1' to every word of second row | 65,568 KB | 44,236 KB | 21,332 KB |

To store 20,000 english words C++ STL Map needs 1,720 KB of memory and the proposed trie needs 1,520 KB of memory. From the above table, we see that the difference of memory consumption is gradually increased when the input size is increased. From the above table, we can see that in case of space complexity the proposed trie is much better than C++ STL Map.

*G. Insertion process of Proposed Trie*

Insertion process of the proposed trie tree is a little bit difficult. There are lots of node breaking and creation process to insert a word to propose trie. For that case, we cannot figure out the exact time complexity of inserting a word in the proposed trie. Here, we show a statistical approach to obtain the time complexity of the insertion process of the proposed trie. We consider the total number of node traversal and loop operation as the total number of operation for inserting a word in the proposed trie.

Table 7: proposed trie insertion analysis

| Input | Description | Total Number of Operation for Inserting this words | Minimum Operations to Insert a single word | Maximum Operations to Insert a single word | Average Operations(Total Operations / Number of Word) |
|---|---|---|---|---|---|
| 20,000 english word | Most common english word | 21,51,712 | 3 | 180 | 107 |
| 4,66,544 english word | All available english word | 2,83,67,327 | 3 | 415 | 60 |

Here we show that the proposed trie needs 21,51,712 operations to store 20,000 english words. On this big data set, the minimum operations need to store a single word is 3 and the maximum operations need to store a single word is 180. On an average 107 number of operations need to insert a single word in the proposed trie tree.

*H. Searching process of Proposed Trie*

Time Complexity of search operation is $O(m * log(n) + l)$. Where '$m$' is the height or maximum number of AVL tree level to reach the corresponding ending node, '$n$' is AVL tree size and '$l$' is word length. Now, we will show an experimental analysis of the time complexity of the searching process of the proposed trie. We consider total number of node traversal and loop operation as the total number of operation for searching a word in our proposed trie tree.

Table 8: proposed trie search analysis

| Input | Description | Searched Word | Total Number of Operation |
|---|---|---|---|
| 20,000 english word | Most common English word | 'mother' | 19 |
| 20,000 english word | Most common English word | 'teacher' | 27 |
| 4,66,544 english word | All available english word | 'mother' | 24 |
| 4,66,544 english word | All available english word | 'teacher' | 35 |

It needs only 19 operations to search the '*mother*' word in proposed trie with 20,000 english words. Proposed trie with 4,66,544 english word, it needs only 24 operations to search the '*mother*' word. Time complexity increases a little bit with the increase of data set size. From the above table, we see that searching time complexity of proposed trie is very low.

*I. Proposed Trie as Data Tank*

Data is nothing but character sequences. A software application's overall space complexity depends on the data structure used on that application. It requires 8 bits to represent an ASCII character cell in memory and 16, 32 or more number of bits to represent a Unicode character cell. In our proposed trie tree we have seen that the character cell requirement to store any kind of data set is very low. We proposed to use the proposed trie as a prime data structure or data tank to any software application and use all other data structure as a non-prime data structure which just stores the node pointer of the proposed trie. This structure is highly memory efficient when use it to design an Operating System.

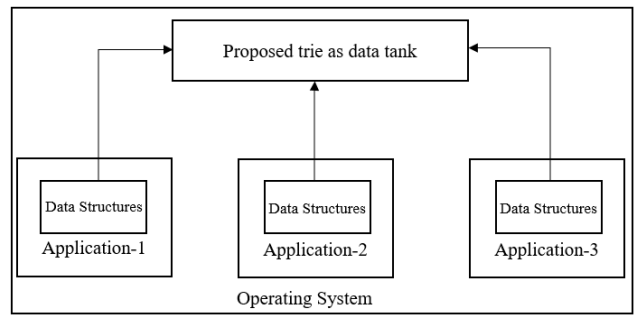

Fig. 11. Proposed trie as data tank (OS)

The above figure all applications data structures point to proposed trie. The application's data structures need not to store character sequences they just need to store a node pointer. Using the above structure, the overall main memory or RAM requirement of a device can be reduced to a great extent. Now we will show an example of this kind of structure. For example, we have the following structure.

| Array1 | Array2 | Array3 |
|---|---|---|
| road | ility | ability |
| abroad | abandon | road |
| ab | ab | ility |

Fig. 12. Example view of word list data

The above structure shows an unintelligent data structure. Array1 first element is equal to Array3 second element. Array1[0] and Array1[1] has the same suffix. Why we use same character sequences or character cell twice or more? This is a wastage of memory. The proposed trie tree has unique emptiness and character-cell reduction capability. Using the proposed trie the above structure becomes.

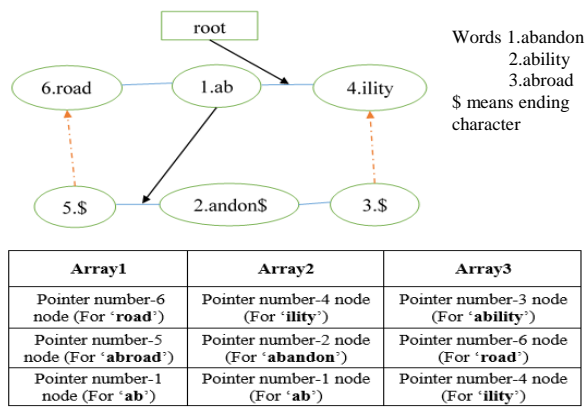

| Array1 | Array2 | Array3 |
|---|---|---|
| Pointer number-6 node (For '**road**') | Pointer number-4 node (For '**ility**') | Pointer number-3 node (For '**ability**') |
| Pointer number-5 node (For '**abroad**') | Pointer number-2 node (For '**abandon**') | Pointer number-6 node (For '**road**') |
| Pointer number-1 node (For '**ab**') | Pointer number-1 node (For '**ab**') | Pointer number-4 node (For '**ility**') |

Fig. 13. Proposed trie as data tank

Here we used the proposed trie as data tank and all other data structure just store a node pointer. Number-5 node (empty node) possesses '*road*' data which inherit from number-6 node. Number-3 node possesses '*ility*' data which inherit from number-4 node. By traversing number-5 node from the root we get '*abroad*' data and traversing number-3 node we get '*ability*' data. In our figure-12 we see Array1[0] and Array[1] have the same suffix '*road*' and it has to store two copy of same data. But using the above structure, we don't need to store the same character sequence more than once. For '*road*' data Array1[0] needs to store the number-6 node pointer. For '*abroad*' data Array1[1] store number-5 node pointer. The above structure doesn't have any data redundancy. Arrays just store a node pointer of the proposed trie. Proposed trie needs a very small amount of character cell. This proposed structure is highly memory efficient. Using proposed trie as data tank any application's space complexity can be reduced to a great extent.

V. Conclusion

We introduce a new emptiness property to radix trie. This trie has prefix, infix and suffix sharing ability. We use a practical approach to prove our algorithm. In the experimental analysis section, we see that the proposed trie is extremely memory efficient than radix trie and some other popular data structures and algorithms. This trie tree has character cell reduction capability and it can dramatically reduce any application's space and time complexity. The proposed trie tree can be used as a data repository of all kinds of data are used in a software application. In our current software development, trie tree is widely used for auto suggestion and spell checking so there's need to store a large word data set or dictionary in a file system. To reduce the app size there requires a reliable file compression algorithm. The character cell reduction property of proposed trie can make him a good dictionary file compression algorithm. In the future, we will publish a file compression algorithm based on the proposed data structure. We believe that the proposed algorithm can fulfill industry needs. Word lookup process is much efficient by using this proposed algorithm.